\documentclass[prl,superscriptaddress,showpacs,reprint]{revtex4-1}

\usepackage{graphicx,textcomp,amssymb,amsmath,dcolumn,hyperref}

\begin{document}

\title{Optical blocking of electron tunneling into a single self-assembled quantum dot}

\author{A.~Kurzmann}
\email{annika.kurzmann@uni-due.de}
\author{B.~Merkel}
\affiliation{Fakult\"at f\"ur Physik and CENIDE, Universit\"at Duisburg-Essen, Lotharstra{\ss}e 1, Duisburg 47048, Germany}
\author{P. A.~Labud}
\author{A.~Ludwig}
\author{A. D.~Wieck}
\affiliation{Chair of Applied Solid State Physics, Ruhr-Universit\"at Bochum, Universit\"atsstr. 150, 44780 Bochum, Germany}

\author{A.~Lorke}
\author{M.~Geller}

\affiliation{Fakult\"at f\"ur Physik and CENIDE, Universit\"at Duisburg-Essen, Lotharstra{\ss}e 1, Duisburg 47048, Germany}

\date{\today}

\begin{abstract}

Time-resolved resonance fluorescence (RF) is used to analyse electron tunneling between a single self-assembled quantum dot (QD) and an electron reservoir. In equilibrium, the RF intensity reflects the average electron occupation of the QD and exhibits a gate voltage dependence that is given by the Fermi distribution in the reservoir. In the time-resolved signal, however, we find that the relaxation rate for electron tunneling is, surprisingly, independent of the occupation in the charge reservoir ---in contrast to results from all-electrical transport measurements. Using a master equation approach, which includes both the electron tunneling and the optical excitation/recombination, we are able to explain the experimental data by optical blocking, which also reduces the electron tunneling rate when the QD is occupied by an exciton.

\end{abstract}

\maketitle

Electron tunneling into semiconductor quantum dots (QDs) has been used to study Coulomb~\cite{warburton1998} and exchange interaction~\cite{petta2005}, as well as to prepare, read-out, and manipulate spin states~\cite{elzerman2004,hanson2005}. It has also been employed to study shot noise~\cite{Gustavsson2006,fricke2007bimodal} and reveal the Fano factor~\cite{Gustavsson2006a} in mesoscopic systems. Most of these transport measurements have been performed on semiconductor QDs, which were defined in a two-dimensional electron gas (2DEG) by lithography techniques~\cite{held1997semiconductor}. Another QD system, which is highly interesting for optical purposes, are epitaxially-grown self-assembled QDs~\cite{Bimberg1998}, where the optical excitonic transitions can be coupled to a photon light field to study quantum optics~\cite{kiraz2004}, e.~g.~in resonance fluorescence~\cite{muller2007, matthiesen2012}. They are also extensively studied for optical devices, like single photon sources~\cite{Michler2000,yuan2002,santori2002}, QD lasers~\cite{saito1996,Ustinov1998} or optical amplifiers~\cite{laemmlin2006}. 

We use here resonance fluorescence (RF) as an optical probe to study for the first time the transport tunneling dynamics between an electron reservoir and a single self-assembled QD. Using voltage pulses and a time-resolved RF detection scheme, we are able to map the Fermi distribution in the electron reservoir and measure the relaxation rate for tunneling between the QD and the charge reservoir. We find clear evidence that the optical excitation of the QD reduces this rate, effectively leading to an optical blocking of single electron tunneling. 

\begin{figure}
  \includegraphics{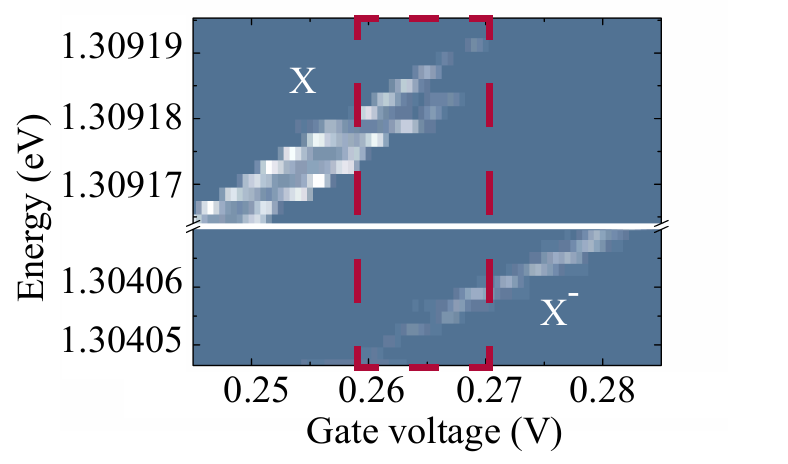}
  \caption{Resonance fluorescence (RF) scan of the exciton ($X$) (with a fine structure splitting of about $8\,\mu\text{eV}$) and trion ($X^-$) for different laser excitation energies and gate voltages.}
  \label{fig1}
  \end{figure}

The investigated sample was grown by molecular beam epitaxy (MBE) and resembles a field-effect-transistor structure~\cite{petroff2001,warburton1998} containing a layer of self-assembled InAs QDs (see also~\cite{supplementary1}).

We use a confocal microscope setup in a bath-cryostat at a temperature of $4.2\,\text{K}$ (see also~\cite{supplementary2}). RF spectroscopy of the exciton $X$ and trion $X^-$ resonances at different gate voltages and laser frequencies shows a transition region between $0.26\,\text{V}$ and $0.27\,\text{V}$ (outlined in red in Fig.~\ref{fig1}), where both transitions are simultaneously visible. Because of the thermally broadened distribution of electrons in the back contact, in this range of gate voltages, the QD is occupied by a single electron with a probability $P$, giving rise to the $X^-$ transition. Correspondingly, the QD will be empty with a probability $1-P$ leading to the observation of the $X$ transition.  

 \begin{figure}
    \includegraphics[scale=1
   ]{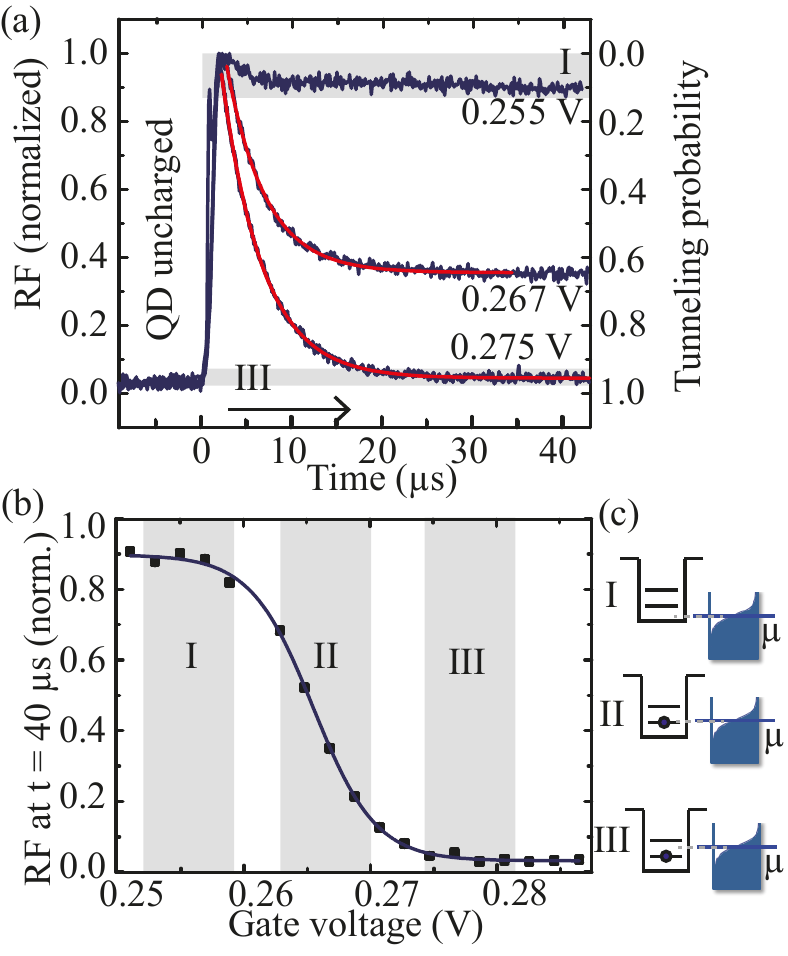}
    \caption{(a) Measured time-resolved RF signal of the exciton transition (trion out of resonance) for three different charging voltages $V_2$ (red lines are exponential fits, used to obtain relaxation rates). (b) Normalized RF signal at $t=0.04\,\text{ms}$ as a function of gate voltage. (c) Schematic representation of the alignment between the Fermi energy in the back contact and the QD state for the three shaded regions in (b). }

    \label{fig2}
    \end{figure}
    
We use a time-resolved RF measurement scheme to investigate the tunneling of a single electron into a single QD by a n-shot measurement~\cite{lu2010}. For each shot, we first prepare an empty QD state by setting the gate voltage to $V_1=0\,\text{V}$, well below the transition region. The laser energy is adjusted so that RF will occur for a gate voltage $V_2$, which lies within or near the transition region. Therefore, as long as $V_1$ is applied to the gate, no RF signal is observed, ---see times $t<0$ in Fig.~\ref{fig2}(a). At $t=0$, the gate voltage is switched to $V_2$, which influences the QD in two ways: (i) On the one hand, it shifts the QD excitonic transition by the quantum confined Stark effect~\cite{Li2000} into resonance with the laser energy and we observe an RF signal by resonant light scattering (RF-signal in Fig.~\ref{fig2}(a) for $t\gtrsim 0\,\text{ms}$). (ii) On the other hand, it shifts the energy levels of the QD with respect to the Fermi distribution in the reservoir. Tunneling can occur when occupied states in the electron reservoir match in energy with empty states in the QD and there will be a non-vanishing probability that the QD will be occupied by one electron. The additional electron switches the exciton transition off, as the transition for a charged QD (the $X^-$ transition) is out of resonance with the laser energy. The evolution from an empty QD (at $t=0$) to a thermal distribution of the QD charge at $t\rightarrow\infty$  is observed as an exponential decay in the RF signal (see Fig.~\ref{fig2}(a)), when averaged over typically $10^6$ voltage pulses.  
    
In Fig.~\ref{fig2}(a), we display the averaged electron tunneling signal on a microsecond time scale for three representative voltages $V_2=0.255\,\text{V}$, $0.267\,\text{V}$ and $0.275\,\text{V}$. The time evolution of the normalized RF signal is nearly constant for $V_2=0.255\,\text{V}$ as no tunneling into the QD is possible (see also panel I in Fig.~\ref{fig2}(c)). For a gate voltage $V_2=0.267\,\text{V}$, an exponential decay of the RF signal is observed that saturates slightly below a value of 0.4. At this voltage, 60 percent of the measurements end in a situation where one additional electron has tunneled into the dot and the $X$ emission quenches (panel II in Fig.~\ref{fig2}(c)). At a gate voltage $V_2=0.275\,\text{V}$, we observe an RF signal of the $X$ only at the beginning of the transient, and it is completely quenched at $t=0.04\,\text{ms}$, indicating that the dot will be occupied by one electron with almost $100\,\%$ probability at this gate voltage.

We changed the voltage $V_2$ in small steps ($2\,\text{mV}$) from  $V_2=0.252\,\text{V}$ to $V_2=0.288\,\text{V}$ and measured the transient of the electron tunneling as discussed above. The black dots in Fig.~\ref{fig2}(b) show  the equilibrium amplitude of the RF signal at $t= 0.04\,\text{ms}$ as a function of the gate voltage. The blue solid line in Fig.~\ref{fig2}(b) is a fit to the data with a Fermi distribution $f(E)$ where temperature, amplitude and chemical potential were taken as free parameters~\cite{Gustavsson2006}. The conversion from gate voltage to energy can be done by $E=e\cdot d_{tunnel}/d_{dot}\cdot V_g=e/\lambda \cdot V_g$, where $\lambda$ is the so-called lever-arm, given by the thickness of the tunneling barrier $d_{tunnel}$ and the distance QD layer to gate contact $d_{dot}$~\cite{warburton1998, drexler1994}. This leads to the lever-arm $\lambda\approx 7 $ in our sample \cite{supplementary1}. The temperature $T=4.2(\pm 0.2)\,\text{K}$ obtained from the fit is in excellent agreement with the base temperature of the helium bath cryostat.

A schematic representation of the three gate voltage regions with low occupation probability (I), occupation probability $\approx 0.5$ (II) and high occupation probability (III) is shown in Fig.~\ref{fig2}(c),---see also grey regions in Fig.~\ref{fig2}(b).

Evaluating the exponential relaxation rates (see red lines in Fig.~\ref{fig2}(a)), we find a constant value of  $\gamma_m=230(\pm 30)\,\text{ms}^{-1}$ over the entire investigated gate voltage range. This observation is quite surprising. A theoretical model developed for transport measurements~\cite{beckel2014b} on similar QDs suggests that $\gamma_m$ should depend on $f(E)$, as discussed in the following. 

Calculations based on a master equation show that the relaxation rate is given by~\cite{beckel2014b,Gustavsson2007a}
\begin{eqnarray}
\gamma_m=\gamma_{Out}+\gamma_{In}, \label{transport}
\end{eqnarray} 
with $\gamma_{In}$ and $\gamma_{Out}$ being the tunneling rates into and out of the QD, respectively. They are given by 
\begin{eqnarray}
\gamma_{In}&=& d_{In}\Gamma f(E)\qquad \text{and} \label{Fermi1}\\
\gamma_{Out}&=&d_{Out}\Gamma(1-f(E)),
\label{Fermi2}
\end{eqnarray}
where $\Gamma$ is the transition rate through the tunneling barrier and $d$ the degeneracy of the final state. In Eq.~\ref{Fermi1}, $d_{In}=2$ to account for the doubly spin degenerate empty QD state. In Eq.~\ref{Fermi2} $d_{Out}=1$ because there is only one possible channel to discharge a singly occupied QD. Hence, $\gamma_m=\Gamma(1+f(E))$ will be dependent on the Fermi function and therefore, on the applied gate voltage.

\begin{figure}
     \includegraphics[scale=1]{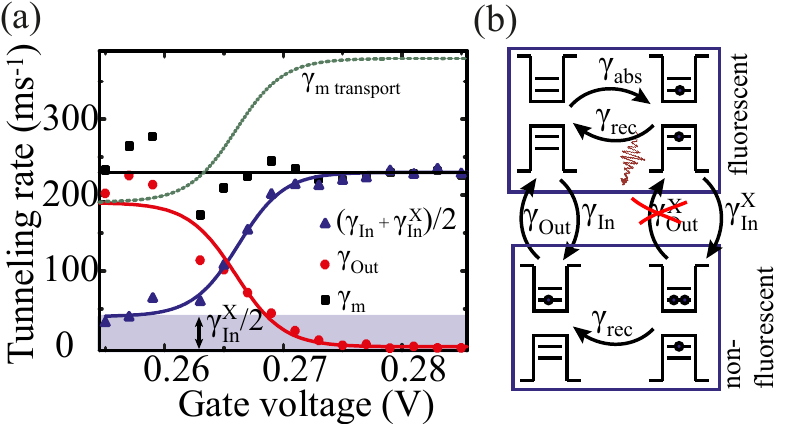}
     \caption{(a) Bare tunneling rates $\gamma_{In}$ and $\gamma_{Out}$ (data points) and fits with Fermi functions (solid lines) versus gate voltage together with the measured relaxation rate $\gamma_m$ (black rectangles) and the transport relaxation rate (green dashed line). (b) Occupation of the fluorescent and non-fluorescent states around $V_2=0.265\,\text{V}$. Arrows indicate optical and transport processes with their respective rates $\gamma$.}
     \label{fig3}
     \end{figure}

We explain the fact that here we do not observe an energy-dependent relaxation time with the influence of the simultaneous optical excitation. To account for the influence of the excitonic state in the QD on the tunneling rates, we extend the master equation approach~\cite{beenakker1991,Gustavsson2007a,beckel2014b} to also include the optical excitation (with absorption rate $\gamma_{abs}$) and recombination (with rate $\gamma_{rec}$) in the QD. We furthermore consider the tunneling rate $\gamma_{In}^X$ of electrons into an exciton state, resulting in a trion~\cite{kloeffel2011}. The reverse process is not possible: the energy of the trion is $\approx 5\,\text{meV}$ smaller than the exciton energy (see~\cite{supplementary3,seidl2005}) so that this process would require tunneling into the back contact well below the Fermi energy, which is Pauli forbidden. Rather we need to consider trion recombination and subsequent tunneling of the remaining single electron with rate $\gamma_{Out}$ (see also arrows in Fig.~\ref{fig3}(b)). We distinguish between fluorescent and non-fluorescent states. The fluorescent state comprises the empty dot and the exciton state, the non-fluorescent state includes the trion as well as the singly charged QD, see upper and lower panel in Fig.~\ref{fig3}(b), respectively.

To solve the Hamiltonian in first order perturbation theory, we make use of the much higher recombination rate $\gamma_{rec}$ compared to the tunneling rates $\gamma_{In/Out}$ (approximately three orders of magnitude). The time evolution of the fluorescent state is then given by the differential equation
\begin{equation}
\dot P_f(t)=-\sigma\gamma_{In}P_{f}(t)+\gamma_{Out}P_{nf}(t) +\gamma_{In}^X (1-\sigma)P_f(t),
\label{equ1}
\end{equation}
where $P_{nf}$ and $P_f$ are the occupation probabilities for the non-fluorescent and the fluorescent state and $(1-\sigma)$ is the average exciton occupation of the QD in the fluorescent state with
\begin{equation}
\sigma=\frac{\gamma_{rec}}{\gamma_{abs}+\gamma_{rec}}.
\end{equation}
By the laser excitation power, $\sigma$ is tunable between $1$ (weak perturbation, i.~e.~no exciton inside the dot) and $0.5$ (saturation, i.~e.~ the QD is occupied by an exciton half of the time). 

The boundary conditions $P_f(0)=1$ and $P_f(t)+P_{nf}(t)=1$ are used to solve Eq.~\ref{equ1}. We obtain
\begin{equation}
P_f(t)=\frac{(\gamma_m-\gamma_{Out})e^{-\gamma_mt}+\gamma_{Out}}{\gamma_m}\,
\label{equ2}
\end{equation}
with the relaxation rate 
\begin{equation}
\gamma_m=\gamma_{Out}+\sigma\gamma_{In}+(1-\sigma)\gamma_{In}^X.
\label{equ3}
\end{equation}
In the experiment, the relative fluorescence amplitude is proportional to the probability that the QD is not charged. Therefore, $P_f(t)$ directly reflects the measured transients in Fig.~\ref{fig2}(a) with a decay constant $\gamma_m$ given by Eq.~\ref{equ3}. The term proportional to $\gamma_{In}^X$ is constant, because this tunneling takes place well below the Fermi energy, where the Fermi function equals 1. 
The remaining two terms in Eq.~\ref{equ3} are similar to the transport relaxation rate in Eq.~\ref{transport}, however, with an additional factor $\sigma$. Thus, the tunable factor $0.5\le\sigma\le 1$ reduces the tunneling rate under illumination (for measurements see supplemental information \cite{supplementary4}). This optical blocking can be understood from the fact that, during the time that an exciton is present in the QD, the number of tunneling paths is reduced from 2 (spin degeneracy of the electron state) to 1. 

For saturation of the $X$ transition ($\sigma=0.5$) this optical blocking compensates the degeneracy factor of 2 in Eq.~\ref{Fermi1} and leads to a relaxation rate $\gamma_m=\Gamma (1-f(E))+0.5\cdot 2\cdot f(E)\Gamma+0.5\cdot \gamma_{In}^X=\Gamma+0.5\cdot \gamma_{in}^X$ that is independent of the Fermi energy. The prediction $\gamma_m=const.$ is in good agreement with our experimental findings (see black data points and line in Fig.~\ref{fig3}(a)) and shows that tunneling between the QD and the back contact can strongly be influenced by simultaneous optical excitation of the QD. For comparison the green dashed line shows the calculated $\gamma_m$ as expected in a pure transport measurement~\cite{Gustavsson2007a}.
 
 Eq.~\ref{equ2} can be used to determine the tunneling rate out of the QD, $\gamma_{Out}$, by the time independent offset (Fig.~\ref{fig2}(b)). The results are plotted in Fig.~\ref{fig3}(a) as red dots together with the tunneling rates into the QD, $0.5\gamma_{In}+0.5\gamma_{In}^X$, which are calculated using Eq.~\ref{equ3} (blue triangles). As mentioned above, $\gamma_{In}^X$ is temperature independent and $\gamma_{in}$ can be fitted with a Fermi function. Using Eq.~\ref{Fermi1} and Eq.~\ref{Fermi2}, we find $\Gamma_{in}^X=80(\pm 20)\,\text{ms}^{-1}$ and $\Gamma=190(\pm10)\,\text{ms}^{-1}$. Thus, the transition rate into the exciton $\Gamma_{in}^X$ is reduced by a factor of 2.4 compared to the transition rate into the empty dot $\Gamma$. This suppression of tunneling can be seen directly in the transients of Fig.~\ref{fig2}. For $V_G=0.255 V$, tunneling into the empty QD is energetically forbidden and the transient only reflects tunneling into the exciton state. We observe a reduction of $P_f$ of only 10 percent and calculate $\Gamma_{in}^X=0.2\cdot\Gamma_{in}$ from Eq.~\ref{equ2}, \ref{equ3}, \ref{Fermi1} and \ref{Fermi2}. In other words, tunneling into an exciton state is strongly reduced compared to tunneling into an empty dot. 
 
At present this additional optical blocking mechanism ($\Gamma_{In}^X\ll \Gamma$) is not fully understood. One possibility would be the energy shift of $5\,\text{meV}$ between the tunneling into the exciton and tunneling into the empty dot. However, a WKB estimate only gives a change of the barrier transparency of roughly 30 percent. The transition rate through the barrier is also dependent on the extent and the orientation of the wave function in the QD, which will be different for an excitonic and empty states. A quantitative estimate of this influence is quite challenging and beyond the scope of this paper.
  \begin{figure}
      \includegraphics[scale=1]{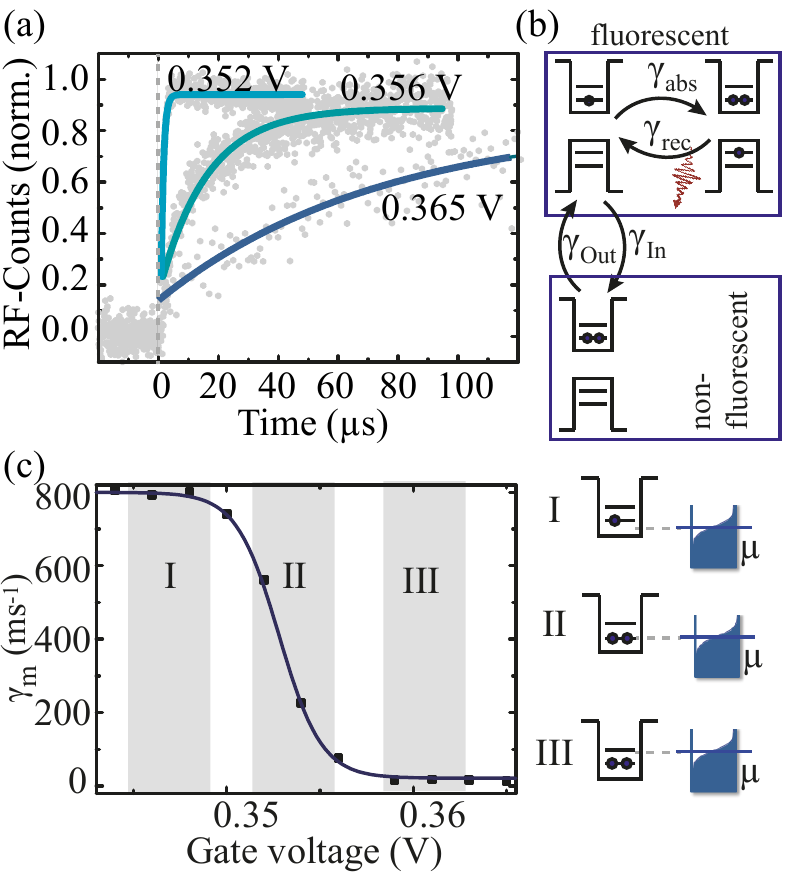}
      \caption{(a) Time-resolved  RF signal from the $X^-$ transition, for different gate voltages when the second electron can tunnel out of the QD. (b) Occupation of the fluorescent and non-fluorescent states around $V_2=0.353\,\text{V}$. Arrows indicate optical and transport processes with their respective rates $\gamma$. (c) Plot of the experimental relaxation rates versus gate voltage with a fitted Fermi function.}
      \label{fig4}
      \end{figure}
      
      In the following we will discuss a tunneling process where the degeneracy and the factor $\sigma$ do not cancel each other. Fig.~\ref{fig4}(a) shows three representative transients in a gate voltage region where a second electron can tunnel into and out of the QD. The measured signal is the RF of the $X^-$ transition. We start with a gate voltage $V_1=0.41\,\text{V}$, where the QD is charged with two electrons and therefore out of resonance with the laser excitation. At $t=0$ we switch to a gate voltage $V_2=0.353\,\text{V}$, where tunneling of one electron out of the QD is possible. At this gate voltage, the $X^-$ resonance matches the laser energy and we observe an increasing RF-signal as the QD reaches equilibrium with the electron reservoir and we have a non-vanishing probability of finding a single electron in the QD, see Fig.~\ref{fig4}(a). The relaxation rates for electron tunneling $\gamma_m$ obtained from the transients are strongly gate voltage dependent as shown in Fig.~\ref{fig4}(c). They decrease from about $800\,\text{ms}^{-1}$ in region I down to almost zero in region III and resemble again the Fermi distribution in the electron reservoir.
      
To explain the gate voltage dependence of $\gamma_m$, we use again the master equation approach with the QD states and rates shown in Fig.~\ref{fig4}(b). We obtain the gate voltage dependent relaxation rate
\begin{equation}
\gamma_m=\Gamma\left(2+\left(\sigma-2\right)f(E)\right),
\label{gm}
\end{equation}
with the degeneracies $d_{In}=1$ and $d_{Out}=2$ for the singly and the doubly charged QD, respectively. For saturated excitation ($\sigma=0.5$), Eq.~\ref{gm} suggests a drop in $\gamma_m$ by a factor of 4 as the Fermi distribution in the back contact is shifted from $f(E)=0$ to $f(E)=1$. Experimentally, however, we find a factor of 60. Therefore, we use $\sigma$ as a fit parameter and find a value of $1/19$, which means that the apparent trion recombination rate is much lower than expected for a transition driven at saturation ($\sigma=0.5$). We explain this results with an Auger-type recombination process~\cite{jang2008enhanced}, that results in an empty QD and lead to a suppressed RF-signal until an electron has tunneled back into the QD from the reservoir. A detailed estimation of the Auger rate is given in the supplemental material~\cite{supplementary5}.

In conclusion, we have investigated the dynamics of electron tunneling between an electron reservoir and a single self-assembled QD under optical excitation. In contrast to transport studies, we find that the relaxation rate is independent of the chemical potential in the back contact. We explain this surprising behavior as a consequence of optical blocking, which also reduces the transition rate into the exciton state. Our findings open up a new route to optically tune the relaxation rate between a QD and a charge reservoir, with a time-resolution, which is only limited by the Rabi frequency of the QD exciton transition.

\end{document}